%% file: 9502043v2.tex
\input preprint
%\Input{figures.aux}
%
\def\eps{\epsilon}
\def\Oeps(#1){{\cal O}(\eps^{#1})}

\def\OZero{{\cal O}(1)}
\mathchardef\aup="0378
\mathchardef\adn="0379
\newbox\bup
\newbox\bdn
\setbox\bup=\hbox{\tenpoint\strut\raise 5pt\hbox{$\aup$}\strut}
\setbox\bdn=\hbox{\tenpoint\strut\raise 5pt\hbox{$\adn$}\strut}
\def\up{{\copy\bup}}
\def\dn{{\copy\bdn}}
\parskip 5pt plus 3pt minus 1pt
\advance\baselineskip+0.2pt
\def\Ref(#1){[#1]}
%
% -----------------------------------------------------------------------------
%
\line{Gen.Rel.Grav. Vol.32(2000) pp.897--918.\hfil}%

\vskip 1cm

\title{%
{Is the Regge Calculus a consistent approximation to}\cr
{General Relativity?}\cr}

\address{%
\cr
{\rm Leo Brewin}\cr
\cr
{\sl Department of Mathematics}\cr
{\sl Monash University}\cr
{\sl Clayton, Vic. 3168}\cr
{\sl Australia}\cr}
%
% Revised  6-May-98, alterations to keep referee's happy
% Revised 17-Jul-98, and again
%

\beginabstract

We will ask the question of whether or not the Regge calculus (and two related
simplicial formulations) is a consistent approximation to General Relativity.
Our criteria will be based on the behaviour of residual errors in the discrete
equations when evaluated on solutions of the Einstein equations. We will show
that for generic simplicial lattices the residual errors can not be used to
distinguish metrics which are solutions of Einstein's equations from those that
are not. We will conclude that either the Regge calculus is an inconsistent
approximation to General Relativity or that it is incorrect to use residual
errors in the discrete equations as a criteria to judge the discrete equations.

\endabstract

\beginsection{Introduction}

Since its inception in 1961 the Regge calculus \Ref(1) has been believed to be a
consistent and convergent approximation to Einstein's theory of General
Relativity. It has often been touted as a natural discretisation of Einstein's
equations and it has also been used as a possible basis for a quantum theory of
gravity (for a detailed bibliography see Williams and Tuckey \Ref(2)). However
the use of the Regge calculus in numerical relativity has been limited to highly
symmetric spacetimes and upon lattices specifically designed for those
spacetimes. Yet little is known about how the Regge calculus performs for
generic spacetimes and it is this class of spacetimes for which the Regge
calculus is most suited.

\input tilde

It is therefore very important that the Regge calculus be tested for
non-symmetric spacetimes and upon generic simplicial lattices. Ideally this
would entail solving the Regge equations for a variety of non-trivial spacetimes
and to compare the solutions against those of the Einstein equations. 
Unfortunately the task of solving the Regge equations for such spacetimes is
way beyond current technology. Thus we are forced to look for some other (less
stringent) criteria.

A much simpler test is to evaluate the residual error in the Regge calculus by
evaluating the Regge equations for a set of leg-lengths computed from a known
solution of Einstein's equations. 

For a typical Regge equation the residual error $\tau_i$ can be defined as 
$$
\tau_i = \sum_{j(i)}\> \theta_j {\partial A_j\over \partial L^2_i}
\eqno\eqndef{TruncErr}
$$
For an exact solution of the Regge equations each $\tau_i$ would be zero.
However, with the leg lengths set from a solution of Einstein's equations one
can not expect the $\tau_i$ to be zero. We would expect, though, that as the
simplicial lattice is refined then the $\tau_i$ should vanish. As we shall
see later, what is most important is not that the $\tau_i$ vanish but how
quickly they vanish with respect to successive refinements in the simplicial
lattice.

We will compute the residual errors for three different simplicial lattices
and for five different smooth metrics (four of which are exact solutions of
Einstein's equations).

The principle result of this paper is that, {\bf the
residual error in the Regge equations cannot be used to distinguish between
metrics which are solutions of Einstein's equations from those that are not}.

This result has been previously noted by Miller \Ref(4), however, we shall
draw sharply different conclusions from those made by Miller.

We shall conclude that either the Regge calculus is not a consistent approximation
to General Relativity or that it is incorrect to use residual errors as
the criteria on which to judge the Regge calculus.

The use of residual errors as a criteria on which to judge a discrete set of
equations is commonly used in the analysis of finite difference approximations
to differential equations. Its value is that it is easy to apply and that quite
often one can show, by analytic methods, that if the residual error vanishes
as the step length is reduced and if the finite difference scheme is stable then
the exact solution of the discrete equations will converge to solutions of the
original differential equations (see for example the Lax Equivalence Theorem
\Ref(5)).

\beginsection{Methods}

We will consider in sequence three main issues. First, the algorithm for
assigning the leg lengths in the lattice. Second, the choice of simplicial
lattices and, finally, the choice of continuum metrics.

Our basic plan is to construct a local group of simplices representing a small
patch on the continuum spacetime and to compute the residual errors in the
discrete equations as a function of the scale of the patch. However, to evaluate
the residual errors one needs to establish some correspondence between the
smooth $g_{\mu\nu}$'s of the continuum and the $L_{ij}$ of the lattice. A
natural way to do this is to map geodesic segments from the continuum to legs of
the lattice. The geodesic length can be computed from the continuum metric and
then assigned to the $L_{ij}$. Thus we are lead to the following two point
boundary problem.

\vskip0.25cm
\testbreak

\begnarrow
Find the curve $x^\mu(\lambda)$ which, for $0\leq\lambda\leq 1$, satisfies
$$0 = {d^2x^\mu\over d\lambda^2} 
     + \sum_{\alpha\beta}\Gamma{}^{\mu}_{\alpha\beta}(x^\rho(\lambda))
                                                     {dx^\alpha\over d\lambda}
                                                     {dx^\beta\over d\lambda}$$
subject to $x^\mu(0)=x^\mu_i$ and $x^\mu(1)=x^\mu_j$.
\endnarrow

\vskip 0.25cm

Once the geodesic is known the squared leg length is assigned as
$$L^2_{ij} = \omega
             \left(\int_0^1\>\big\vert g_{\mu\nu}
                  {dx^\mu\over d\lambda}
                  {dx^\nu\over d\lambda}\big\vert^{1/2}
                                             \>d\lambda\right)^2$$
with $\omega=\pm1$ according to the signature of the geodesic (which
is known to be constant along the geodesic).

This boundary value problem was solved by a shooting method. The idea is to
convert the boundary value problem into an initial value problem and then to try
to find the initial values so as to satisfy the end boundary condition. Let
$y^\mu(\lambda,g^\nu)$ denote a solution of the initial value problem starting
with the guess $g^\mu=dy^\mu/d\lambda$. The strategy now is to solve the
coupled equations $0=x^\mu(1)-y^\mu(1,g^\nu)$ for $g^\nu$. This was done via a
Newton-Raphson approach
$$\eqalign{g^\mu_{n+1} &= g^\mu_n + \delta g^\mu\cr
           x^\mu(1)-y^\mu(1,g^\nu_n) &= - \left(
                                        {\partial y^\mu\over\partial g^\alpha}
                                        \right)_n
                                        \delta g^\alpha\cr}
                                        $$
The partial derivatives were evaluated numerically
$$\left({\partial y^\mu\over\partial g^\alpha}\right)_n =
{y^\mu(1,g^\nu_n+\delta h^\nu)-y^\mu(1,g^\nu_n-\delta h^\nu)
 \over 2\delta h^\alpha}
\eqno\eqndef{Partials}
$$
Notice that to evaluate all of the partial derivatives requires at least 32
complete integrations of the initial value problem. The leg lengths were
evaluated by appending the differential equation for $ds/d\lambda$ to the
Runge-Kutta routine.

We will need to compute the residual errors for series of patches of varying
sizes. This is actually rather easy to do. Let a set of coordinates in the
patch on the continuum spacetime be given by $y^\mu$ and their values at the
vertices by $y^\mu_i$. Now define a coordinate transformation from $y^\mu$
to $x^\mu$ by
$$
y^\mu = x^\mu_{\star} + \eps(x^\mu-x^\mu_{\star})
\rlap{$\displaystyle\hskip3cm 0<\eps\leq1$}
$$
This will be viewed as an active transformation having the effect of
focusing the vertices upon the (freely chosen) central point $x^\mu_{\star}$.
In this construction each vertex carries with it its initial coordinates.
The metric in these coordinates is therefore
$$
ds^2 = \eps^2 g_{\mu\nu}(x^\mu_{\star} + \eps(x^\mu-x^\mu_{\star}))
                dx^\mu dx^\nu
$$
The leading $\eps^2$ serves only as a constant conformal factor and can thus
be absorbed by a re-scaling $ds\leftarrow ds/\eps$. Finally, the form of the
metric from which the geodesics are calculated is just
$$
ds^2 = g_{\mu\nu}(x^\mu_{\star} + \eps(x^\mu-x^\mu_{\star}))
       dx^\mu dx^\nu
\eqno\eqndef{dsSquare}
$$
with the coordinates at the vertices $x^\mu_i$ being chosen independently of
$\eps$.

Let us now turn to the choice of simplicial lattices. We will consider just three
distinct models. Two of the models were constructed from a set of tetrahedra
surrounding a common vertex with the 4-simplices being generated by dragging the
common vertex forwards and backwards in time. The third model was a 2x2x2x2 four
dimensional hypercubic lattice. The details of these models are as follows.

\beginsubsection{Model 1}

This model is based on the collection of four tetrahedra described by
$${%
\def\g{\hskip15pt}(1,2,3,5)\g (1,2,4,5)\g (1,3,4,5)\g (2,3,4,5)}$$
as depicted in Figure (\figrfr{ModelOne}). Spatial coordinates are then assigned
to each of the five vertices
$$(i) \leftarrow (0,x^2_i,x^3_i,x^4_i).$$
Next, a pair of new vertices $(5^\up)$ and $(5^\dn)$ are created with
coordinates
$$\eqalign{%
(5^\up) &\leftarrow (+x^1_5,x^2_5,x^3_5,x^4_5)\cr
(5^\dn) &\leftarrow (-x^1_5,x^2_5,x^3_5,x^4_5)\cr}$$
Finally the set of 4-simplices is created by connecting $(5^\up)$ and $(5^\dn)$
to all of the vertices $(1),(2),(3),(4)$ and $(5)$. This creates
a simplicial lattice with eight 4-simplices
$${%
\def\g{\hskip15pt}\eqalign{%
(1,2,3,5,5^\up)\g (1,2,4,5,5^\up)\g (1,3,4,5,5^\up)\g (2,3,4,5,5^\up)\cr
(1,2,3,5,5^\dn)\g (1,2,4,5,5^\dn)\g (1,3,4,5,5^\dn)\g (2,3,4,5,5^\dn)\cr}}$$

\beginsubsection{Model 2}

This is identical in construction to the previous model with the exception that
the initial set of tetrahedra, depicted in Figure (\figrfr{ModelTwo}), consists
of eight tetrahedra
$${%
\def\g{\hskip15pt}\eqalign{%
(1,2,4,5)\g (1,2,5,6)\g (1,2,6,7)\g (1,2,4,7)\cr
(1,3,4,5)\g (1,3,5,6)\g (1,3,6,7)\g (1,3,4,7)\cr}}$$
In this model the common vertex is $(1)$ and it is dragged forward and backward
in time just as in the previous model.

\beginsubsection{Model 3}

This is a four dimensional hypercubic lattice containing 384 4-simplices.
This model can be built from a template hypercube, consisting of 24 4-simplices,
replicated once along each of the four dimensions of the lattice.
The details are as follows.

A single hypercube can be defined recursively as follows. Starting with one
leg $(0,1)$ apply the following rules
$$
\eqalign{%
(a,b) &\mapsto (0a,0b,1b) + (0a,1a,1b)\cr
(a,b,c) &\mapsto (0a,0b,0c,1c) + (0a,0b,1b,1c) + (0a,1a,1b,1c)\cr
(a,b,c,d) &\mapsto (0a,0b,0c,0d,1d)
                  +(0a,0b,0c,1c,1d)
                  +(0a,0b,1b,1c,1d)
                  +(0a,1a,1b,1c,1d)\cr}
$$
The notation $ia$ means append $i$ as the most significant bit in the binary
representation of $a$. The first rule generates two triangles from each leg,
the second, three tetrahedra from each triangle and the final, four 4-simplices
from each tetrahedron. Thus from one leg we will obtain 24 4-simplices defined
by 16 vertices.

The above rules naturally assign a binary number to each of the 16 vertices.
The bits of the binary number can be thought of as coordinates of the vertex
(ie. the vertex with label $(abcd)$ is located at the point with coordinates
$(a,b,c,d)$ and $a,b,c,d$ are in the set $\{0,1\}$). Note that these coordinates
have nothing to do with metric or the physical coordinates of the lattice.
They are introduced solely as an aid in the construction of the lattice.

This hypercube was used as a template in constructing a 2x2x2x2 hypercubic
lattice with the 81 vertices labelled in lexicographic order. For the vertex
with coordinates $(a,b,c,d)$, where $a,b,c,d$ are drawn from the set
$\{0,1,2\}$, the lexicographic label is defined by
$${\tt lex}(a,b,c,d) =  d + 3( c +3( b +3a ) )$$
If we take $(p,q,r,s,t)_i$ to be $(p+i,q+i,r+i,s+i,t+i)$ then the full set of
384 4-simplices of the 2x2x2x2 hypercubic lattice are given by
$${%
\def\s{\phantom{1}}%
\def\g{\hskip15pt}\eqalign{%
(   0,\s 1,\s 4,  13,  40)_i\g 
(   0,\s 1,\s 4,  31,  40)_i\g 
(   0,\s 1,  10,  13,  40)_i\g 
(   0,\s 1,  10,  37,  40)_i\cr 
(   0,\s 1,  28,  31,  40)_i\g 
(   0,\s 1,  28,  37,  40)_i\g 
(   0,\s 3,\s 4,  13,  40)_i\g 
(   0,\s 3,\s 4,  31,  40)_i\cr 
(   0,\s 3,  12,  13,  40)_i\g 
(   0,\s 3,  12,  39,  40)_i\g 
(   0,\s 3,  30,  31,  40)_i\g 
(   0,\s 3,  30,  39,  40)_i\cr 
(   0,\s 9,  10,  13,  40)_i\g 
(   0,\s 9,  10,  37,  40)_i\g 
(   0,\s 9,  12,  13,  40)_i\g 
(   0,\s 9,  12,  39,  40)_i\cr 
(   0,\s 9,  36,  37,  40)_i\g 
(   0,\s 9,  36,  39,  40)_i\g 
(   0,  27,  28,  31,  40)_i\g 
(   0,  27,  28,  37,  40)_i\cr 
(   0,  27,  30,  31,  40)_i\g 
(   0,  27,  30,  39,  40)_i\g 
(   0,  27,  36,  37,  40)_i\g 
(   0,  27,  36,  39,  40)_i\cr}}
$$
for $0\leq i\leq {\tt lex}(1,1,1,1)$.

The artificial coordinates $(a,b,c,d)$ on each vertex can now be replaced with
those appropriate for the particular spacetime under consideration.

The coordinates for the vertices for the first two models models were chosen
according to Table (\tblrfr{VerticesTable}) while for the hypercubic lattice
the vertex labeled ${\tt lex}(a,b,c,d)$ was assigned the coordinates
$$
x^\mu(a,b,c,d) = {1\over4}\left( a\delta^\mu_1
                                +2(b+1)\delta^\mu_2
                                +2(c+1)\delta^\mu_3
                                +2(d+1)\delta^\mu_4 \right)
$$
with $x^\mu_\star = x^\mu(1,1,1,1)$.

Four exact solutions of the vacuum Einstein equations were used in evaluating
the residual errors.

\beginsubsection{Metric 1 : Schwarzschild}

The metric is described, in isotropic coordinates, by
$$
  ds^2 = -f^2(x,y,z)dt^2 + g^2(x,y,z)(dx^2+dy^2+dz^2)
$$
where
$$
  f(x,y,z) = {\left(2r-m\over 2r+m\right)}\>,\hskip 3cm
  g(x,y,z) = \left(1+{m\over2r}\right)^2
$$
and $r^2 = x^2+y^2+z^2$ and the mass was set at $m=1$.

\beginsubsection{Metric 2 : Kasner}

This solution is described, again in pseudo-Cartesian coordinates, by
$$
ds^2 = - e^{2t}dt^2 + e^{4t/3}dx^2 + e^{4t/3}dy^2 + e^{-2t/3} dz^2
$$
It was chosen because it is asymmetrical and thus provides a more
demanding test than that given by the Schwarzschild metric.

\beginsubsection{Metric 3: Plane wave}

This is one of the many exact plane wave solutions and is described by
$$
ds^2 = -4dudv + dx^2 +dy^2 + \left( (x^2-y^2)\sin u -2xy\cos u \right) du^2
$$
with $u=(t-x)/2,\>v=(t+x)/2$.

\beginsubsection{Metric 4: Plane symmetric wave}

A second gravitational wave solution is the plane symmetric wave in the
form
$$
ds^2 = {1\over\sqrt{z}}\left(-dt^2 + dz^2\right) + z\left(dx^2 + dy^2\right)
$$

\beginsubsection{Metric 5: Reference}

If we believe that the Regge calculus is a consistent approximation to General
Relativity then we could reasonably expect the residual error to behave
differently for exact solutions of Einstein's equations than for non-solutions.

Thus we chose to include a metric which was clearly not a solution of
the vacuum Einstein's equations, namely,
$$
ds^2 = -f^2(x,y,z)dt^2 + (dx^2+dy^2+dz^2)
$$
with $f(x,y,z)$ chosen as in the Schwarzschild metric. We will use this
metric as a reference by which we will compare the residual
errors for the first four metrics.

For each of the above metrics and for each of the models the geodesics
were computed with a step length in the Runge-Kutta routine of
$\Delta\lambda = 1/10$ and ten iterations of the Newton-Raphson scheme. These
figures were arrived at by experimentation. Increasing the number of iterations
or decreasing $\Delta\lambda$ seemed to have no significant effect on the
residual errors. In evaluating the partial derivatives (\eqnrfr{Partials}) 
the $\delta h^\mu$'s were chosen as the constant value
$\max_\nu(x^\nu(\lambda=1)-x^\nu(\lambda=0))/50$. Again, this was
arrived at by experimentation (almost identical results were obtained using
$\Delta\lambda=1/5$, five iterations in the Newton-Raphson scheme and
$\delta h=\max_\nu(x^\nu(\lambda=1)-x^\nu(\lambda=0))/25$).

\beginsection{The lattice equations}

We have computed the residual errors for three different set of lattice
equations. The most familiar is that of the Regge calculus. The other two are
variations which though they have a Regge feel to them are in fact distinct from
the Regge calculus.

\beginsubsection{The Regge Calculus}

The standard Regge equations are
$$
0 = \sum_{j(i)}\> \theta_j {\partial A_j\over \partial L^2_i}
\eqno\eqndef{ReggeEqtn}
$$
where the sum includes all of the triangles attached to a leg. There is one
such equation for each leg in the lattice.

\beginsubsection{Miller's equations}\secdef{MillersEqtns}

Miller \Ref(4) obtained a set of equations by taking variations of
the Hilbert action with respect to $g_{\mu\nu}$ rather than the $L_{ij}$.
His equations are
$$
0 = \sum_i\> \left(m\Delta x^\mu\Delta x^\nu\right)_i
    \sum_{j(i)}\>\theta_j {\partial A_j\over \partial L^2_i}
\eqno\eqndef{MillerEqtn}
$$
where the outer sum is a sum over a set of legs, $\Delta x^\mu$ is the vector
joining tip to tail of the leg and $0\leq m\leq1$ is a weighting assigned to the
leg. The second sum is exactly the Regge equation for this leg (and thus
Miller's equations are linear combinations of the Regge equations). The $\Delta
x^\mu$ are computed with respect to a locally flat background metric (the error
in doing so is of the order of a typical defect angle and these vary as
$\Oeps(2)$). There are two issues that remain, first, which legs should appear
in the outer sum and, second, how should the weights $m$ be calculated. Miller
applied his equations only to a single hypercube within a 3x3x3x3 hypercubic
lattice (ie. the hypercube built from the vertex ${\tt lex}(1,1,1,1)$). He also
defined the weights to be the fraction of the leg shared with adjacent
hypercubes. Thus $m=1$ for legs solely within the central hypercube and $m=1/n$
for legs shared by the central and $n-1$ adjacent hypercubes.

We are unable to apply Miller's rules to both of our first and second models for
it is not clear how to compute the Regge equations on the boundary legs nor how
to assign a weight for such legs. Thus we shall choose to include, in the outer
sum, only those legs attached to the central vertex of each model (ie. vertices
5,1 and ${\tt lex}(1,1,1,1)$ for models 1,2 and 3). This is equivalent to
setting $m=1$ for legs directly attached to the central vertex and $m=0$ for all
other legs.

\beginsubsection{Brewin's equations}

Brewin \Ref(6) has proposed the following lattice equations
$$0=\sum_j
    \sum_{i(j)}\>\left(\Delta x^\mu \Delta x^\nu\right)_i
                 \theta_j {\partial A_j\over\partial L^2_i}
\eqno\eqndef{BrewinEqtn}
$$
The outer sum contains only those triangles attached to the central vertex while
the inner sum is a sum over the three legs of each triangle. There is one set of
such equations for every vertex in the simplicial lattice. Note that these
equations can be written as linear combinations of the Regge equations plus
non-zero contributions from legs not attached to the central vertex.

These equations were derived by an integration of the vacuum Einstein equations
directly on the lattice. The techniques used are very similar to those commonly
used in finite element methods and amounts to nothing more than repeated
integration by parts (for the details see \Ref(6)).

Note that if it were not for the weights and the surface terms, equation
(\eqnrfr{MillerEqtn}) would be identical to equation (\eqnrfr{BrewinEqtn}).

\beginsection{Results}

We are primarily interested in the behaviour of the residual error $\tau$ as a
function of the scale parameter $\eps$, though we know that $\tau$ will also
depend on the choice of metric, the lattice and the coordinates assigned to each
vertex.

We shall begin by first forming some simple estimates for the behaviour of $\tau$
for small values $\eps$. 

First we note that if each bone in the lattice is non-null then $\tau$ is a
smooth function of $\eps$. This follows from the simple observation that a term
in $\sum_i \theta_i \partial A_i/\partial L_j$ is singular only when its
corresponding bone is null (in which case the defect is undefined). However by a
careful choice of lattice and coordinates for each vertex we can always find an
$\eps'$ such that for each $0<\eps<\eps'$ each bone has a fixed signature and
consequently each term in the residual error is a bounded function of
$\eps$.

We can also see that $\lim_{\eps\rightarrow 0} \theta_i = 0$ while
$\lim_{\eps\rightarrow 0} \partial A_i/\partial L_j$ remains finite but non-zero.
These deductions follow by inspection of the metric in (\eqnrfr{dsSquare}).
Clearly this metric
is flat in the limit as $\eps\rightarrow 0$ and thus the defects vanish.
Consequently we must have
$$
\tau_i(g,\eps) = \eps^p Q(g,\eps)
$$
where $p$ is some undetermined (positive) number and $Q(g,\eps)$ is some undetermined
function of $\eps$. All that we need to know about $Q(q,\eps)$ is that there
exists numbers $m(g)$ and $M(g)$ such that
$0 < m(g) < \vert Q(g,\eps)\vert < M(g)$ for all $\eps$ in $(0,\eps')$.
That $\vert Q(g,\eps)\vert < M(g)$ follows immediately from the fact that
$\tau_i$ is bounded from above. That $0 < m(g)$ follows
by choosing $p$ such that $\lim_{\eps\rightarrow 0}\tau_i(g,\eps)\eps^{-p}$
remains finite and non-zero.

Note that $p,m,M$ will depend not only upon the metric but also upon the
topology of the lattice and the coordinates assigned to each vertex. They are
however independent of $\eps$.

Consider now a second metric (eg. the reference metric) for which we have
$$
\tau_i(g',\eps) = \eps^q Q(g',\eps)
$$
in the interval $0<\eps<\eps''$ and $q$ is some number (which may differ
from $p$).

Our aim is to compare the residual errors for a pair of metrics, only one
of which is a  solution of Einstein's equations. Thus we are not particularly
interested in the actual values of $p$ and $q$ but rather the value of $p-q$
(though later we will argue that $q=2$ for all choices of $g'$).

Thus consider
$$
{\tau_i(g,\eps)\over\tau_i(g',\eps)} = \eps^{p-q}\left({Q(g,\eps)\over
                                                        Q(g',\eps)}\right)
$$
for $0<\eps<\eps^{\star}$ and $\eps^{\star}={\rm min}(\eps',\eps'')$.
Since $0 < m(g')$, the ratio $Q(g,\eps)/Q(g',\eps)$ is bounded.
Thus%
\footnote{${}^\dagger$}{By $\tau=\Oeps(p)$ we mean that there exists a
constant $K$
and $\eps^{\star}$ such that $\vert\tau\vert \leq K \eps^p$ whenever
$0<\eps< \eps^{\star}$, see \Ref(11)}
not only do we have $\tau_i(g,\eps)=\Oeps(p)$ and
$\tau_i(g',\eps) = \Oeps(q)$ but we also have
$\tau_i(g,\eps)/\tau_i(g',\eps) = \Oeps(p-q)$.

Note that these estimates are valid only when the signature of each bone
is independent of $\eps$ for $0<\eps<\eps^{\star}$. In every one of our cases
studies this condition was met with $\eps^{\star}=0.5$.

For each model and for each metric the {\bf effective residual
errors} were defined as
$$
\eta\strut_2(g,g')  = \left({\sum_i \tau^2_i(g,\eps)\over
                             \sum_i \tau^2_i(g',\eps)}\right)^{1/2}
$$
where $g'$ denotes the reference metric and the sums include every Regge
equation in the computational patch.

If the Regge calculus is a consistent approximation to General Relativity then
we should see $\eta_2(g,g') =\Oeps(r)$ with $r>0$. On the other hand if we observe
$\eta_2(g,g') = \OZero$ then we have some explaining to do for in that case we
are unable, by these means, to distinguish between metrics which are solutions
of Einstein's equations from those that are not.

The results for each of the numerical experiments are displayed in Figures
(\figrfr{FigD}--\figrfr{FigL}). However, their interpretation is not completely
clear cut. In some cases, such as in Brewin's and Miller's equations with Model
3, see Figures (\figrfr{FigI}--\figrfr{FigL}), we can see that effective
residual errors for the exact solutions (metrics 1-4) vary as $\Oeps(2)$
compared to the $\OZero$ for the reference metric. This is what we would expect
for a consistent discrete approximation to Einstein's equations. Yet for other
cases, such as for those of Model 1,
see Figures (\figrfr{FigD},\figrfr{FigG},\figrfr{FigJ}), we
see no distinction between the effective residual errors for the five metrics. 

The worst results were found for Model 1 where the effective residual
errors varied as $\OZero$ for every metric and for each set of lattice
equations. Models 2 and 3 displayed some degree of success but this may be
attributed to their specialised construction (their legs were
aligned to the coordinates axes of the continuum metric and the vertices were
symmetrically placed relative to the central vertex). 

To test this view we decided to repeat the calculations after introducing small
fluctuations in the coordinates of the vertices
$$
x^\mu_i \leftarrow x^\mu_i + 0.05(\omega_i-0.5) x^\mu_i
$$
where $0\leq\omega_i\leq1$ was a random number. By making these small changes we
believe that we are using a simplicial lattice much closer to what we can expect
in a generic application of the Regge calculus (or some other set of equations).
Though these are, however, only minor changes.

We then found, without exception, that the effective residual errors varied as
$\OZero$. Thus we conclude that, for a generic simplicial lattice, the
residual errors in each of the discrete equations
(\eqnrfr{ReggeEqtn}--\eqnrfr{BrewinEqtn}) can not be used to distinguish between
metrics which are solutions of Einstein's equations from those that are not. 

\beginsection{Discussion}

It would seem from the above that we are forced to accept one of two options :
either that the Regge equations are not a consistent discretisation of
Einstein's equations or that it is incorrect to use the residual errors of the
Regge equations as a criteria in this context.

A common objection to this line of reasoning is that the negative result, that
the effective residual varied as $\OZero$, is a consequence of choosing either
an inappropriate lattice, reference metric or vertex coordinates, and that with
a better choice one might observe $\Oeps(2)$ variation in the effective residual. 
We can, however, quickly exclude the option of changing reference metrics in
the hope of improving the convergence. To see this
recall that
$$
\sum_i \theta_i A_i =
\int_{M}\> \eps^2 R(x^\mu_{\star} + \eps(x^\mu-x^\mu_{\star}))\>dV
$$
where the sum includes all bones inside the computational cell $M$ and
$\eps^2 R$ is the scalar curvature of the conformal metric (\eqnrfr{dsSquare}).
Over the patch $M$ the curvature is bounded and the limits of integration
do not depend on $\eps$, thus
$$
\sum_i \theta_i A_i = \Oeps(2)
$$
and as both $A_i$ and $L^2_j$ are $\OZero$ we find that
$$
\sum_i \theta_i {\partial A_i\over \partial L^2_j} = \Oeps(2)
$$
That is, for any choice of reference metric, we can expect
$\tau_i(g',\eps) = \Oeps(2)$ (this same statement has been made by Miller, see
\Ref(4)) and consequently the effective residual $\eta_2(g,g')$ is independent
of the choice of the reference metric.

Thus improving the convergence of $\eta_2$ for a given solution of Einstein's
equations can only be achieved by changing the structure of the lattice. This
has already been demonstrated with Models 1 and 3 for the Kasner metric where
the convergence was seen to be $\OZero$ and $\Oeps(2)$ respectively. However the
real point is that there do exist cases where the effective residual does vary
as $\OZero$. In this specific case, with this lattice, metric etc.  the
effective residual makes no distinction between this solution of Einstein's
equations and the reference metric. How can we explain this behaviour? Or do we
have to accept that there are some classes of lattices for which the method of
effective residuals is inappropriate? If so, then how do we characterise such
lattices? We know of no way to do so and that is why we have included the second
of our two options in the opening paragraph of this section.

If we reject the use of residual errors then we must justify why we can reject
such a useful technique and furthermore we would be required to propose some
other suitable criteria. One could well argue that what really matters is the
convergence (or otherwise) of the solutions of the Regge equations to solutions
of Einstein's equations. If we find convergence to any desired solution of
Einstein's equations then the above behaviour of the residual errors just
becomes a curious fact and the Regge calculus lives to fight another day.

Thus it seems reasonable to speculate on how one might achieve convergence at the
level of solutions but non-convergence at the level of the field equations.

Consider, as a toy example, a second order differential equation
$$
0 = {\cal L}(y)
$$
for some function $y(x)$. Now construct a new function
$${\Tilde y}_\eps(x) = y(x) + \eps^2 f(x/\eps)$$
where $\eps$ is arbitrary and $f(x)$ is any smooth bounded function of $x$. The
function ${\Tilde y}_\eps$ will be the solution of some other second order
equation
$$
0 = {\Tilde{\cal L}}_\eps({\Tilde y}_\eps)
\eqno\eqndef{AlteredEpsF}
$$
Then it is easy to see that
$$\openup3pt\displaylines{%
\vert y - {\Tilde y}\vert = {\cal O}(\eps^2)\cr
\noalign{while}
\vert {\cal L}(y) - {\Tilde{\cal L}}_\eps(y)\vert = {\cal O}(1)\cr}
$$
where $\vert\cdots\vert$ is a point norm.

Thus the residual error is of ${\cal O}(1)$ yet the solutions are second order
convergent. Thus in this toy model we can see that measuring the residual
error alone is not sufficient to declare the discrete scheme to be invalid.

Even if we accept this, there will be some strange behaviour in the equations and
their solutions. The term $\eps^2 f(x/\eps)$ corresponds to some irregular wave
(since $f(x)$ is bounded) on the solution. The amplitude will decrease with
$\eps$ but the frequency will increase. Thus this behaviour could be spotted by
the appearance of a high frequency signal in the solution. Recent results by
Gentle and Miller \Ref(3) displayed exactly this behaviour though it is not yet
known if it is related to the mechanism described here. In the field equations
this term $\eps^2 f(x/\eps)$ appears as a term of the form $f''(x/\eps)$. This
also represents a high frequency term. Note that there is no well defined limit
for this term as $\eps\rightarrow0$ and thus we obtain a differential equation which
does not have a well defined form.

If we are prepared to accept this mechanism then we must somehow explain why
it has not been observed in previous applications of the Regge calculus.

All previous calculations with the Regge calculus (with the exception of the
perihelion calculations of Brewin \Ref(7)) have been consistent with what one
would expect from the Einstein equations. However, without exception, every
numerical spacetime constructed to date using the Regge calculus has possessed a
high degree of symmetry and those symmetries have been explicitly included in
the simplicial lattice. Thus it is arguable that the Regge equations, for those
spacetimes, were of such a simple form that they had no option but to converge
to the correct Einstein equations. Indeed both the Kasner (Lewis \Ref(8)) and
Schwarzschild (Wong \Ref(9)) spacetimes have been successfully constructed from
the Regge calculus, a result which could not be expected nor discounted from the
results presented here. One difference between their work and ours is their
choice of lattice. Where we chose simplices they chose blocks specifically
designed to reflect the symmetries of their target spacetimes. This might be
significant if there is a non-uniqueness in the limiting form of the Regge
equations. This is a real possibility when one considers the infinite choices
one has in sub-dividing spacetimes into 4-simplices. It is also conceivable that
for a given sequence of sub-divisions, the nature by which the leg-lengths are
adjusted as the continuum is approached may have a bearing on the limiting form
of the equations (eg. how are limiting spacetimes with a fractal-like structure
avoided?).

One of the common arguments used to support the view that the Regge calculus
should be a valid approximation to Einstein's theory is that both theories are
derived from the same Hilbert action. However this similarity may not be as
strong as it seems. The important point is that the configuration space for
simplicial metrics is larger than that for smooth metrics (ie. metrics with
bounded curvatures). To see this notice that every smooth metric can be
arbitrarily approximated by a sequence of discrete metrics yet it is not possible
to do the converse, to approximate a given discrete metric by a sequence of
metrics with bounded curvatures. Thus the Regge equations arise from variations
of the metric over a wider class than used in deriving the Einstein equations.
Consequently one can not expect the two sets of equations to agree.

However, one can expect that the Einstein equations would arise as some linear
combination of the Regge equations (arising from the constraints imposed in the
reduction of the configuration space). In fact we have already seen linear
combinations of the Regge equations arising naturally in both Brewin's and
Miller's alternative theories (equations (\eqnrfr{BrewinEqtn}) and
(\eqnrfr{MillerEqtn})). This averaging process may well also wash out the
high frequency term $f''(x/\eps)$ noted above.

A simple example of this is given by the action
$$I(\theta,\phi) = \int \left( \theta^2_{x} + \phi^2_{y} \right)\>dxdy$$
where $\theta$ and $\phi$ are arbitrary smooth functions of $(x,y)$. The 
configuration space is the set of all pairs of smooth functions of two variables.
Extremisation of $I$ over this configuration space leads to two equations
$$
0 = \theta_{xx}\hskip2cm
0 = \phi_{yy}
$$
while extremisation over the reduced configuration space where $\theta = \phi$
leads to the one equation
$$
0 = \phi_{xx}+\phi_{yy}
$$
Solutions of the original pair of equations are always solutions of the later
equation. But the converse is not true, there exist many solutions of Laplace's
equation which are not solutions of the previous pair. If this carries over
to the Regge calculus then we can expect that Einstein's equations will be
recovered as a linear combination of the Regge equations. Thus we can see
that there may exist solutions of Einstein's equations which will not be
solutions of the individual Regge equations though they may be solutions
of an appropriately chosen linear combination of the Regge equations.

This observation addresses, in part, the claims of Barrett \Ref(20)
that solutions of the linearized Regge equations should converge, in the sense
of distributions, to solutions of the Einstein equations subject to a reasonable
bound on the defects. However Barrett does not make any claims about the converse
process, solutions of Einstein's equations being solutions of Regge's equations.

For these reasons we do not believe that the Regge equations (\eqnrfr{ReggeEqtn})
can be used to obtain accurate and consistent approximations to solutions of
Einstein's equations for generic spacetimes. This claim is of course speculative
and is open to criticism.

Let us now return to the question of finding a suitable criteria on which to
judge the consistency of any proposed discrete set of equations. One option is
to compare solutions of the discrete and exact equations. Yet, this is very
difficult to do and is rather time consuming (we need to solve both set of
equations -- why do a job twice?).

We do however have an alternative in which we compute the residual error in the
Einstein equations when evaluated on solutions of the discrete equations. This
must surely be a stronger test than what we have been using. For if the
residual error behaves appropriately (ie. vanishes at an appropriate rate)
then there can be no doubt that the discrete equations do yield valid
approximations to Einstein's equations and that the discrete solutions are
correct.

How would we perform such a computation? It would require us to extract a smooth
metric and in particular a point estimate of the curvature tensor (or Ricci
tensor at worst) from the leg lengths of the lattice. To do so we will need to
surrender the piecewise flat assumption (otherwise we do not get point estimates
for the curvatures). Instead we could to try to fit a locally quadratic expansion
of a smooth metric to the data on the lattice. The quadratic terms in this
expansion should then give us our point estimate of the curvatures. This is very
easily done if one uses Riemann normal coordinates and if one assumes the legs
of the lattice to be short geodesic segments. 

Notice also that in this approach there is actually no need for a separate set of
discrete equations since we are able to use the Einstein equations directly on
the lattice.

This work is currently in progress and we shall report on this in a later paper.

\beginsection{Acknowledgements}

An earlier version of this paper appeared on the Los Alamos National Laboratories
preprint server as {\bf gr-qc/9501023}.

\bgroup
\beginsection{References}
\vskip 0.5cm
\parskip=0pt plus 0pt minus 0pt

\R 1!Regge, Tullio. (1961)!
     General Relativity without Coordinates.!
     Il Nuovo Cimento. Vol.XIX,No.3(1961) pp.558-571.\par

\R 2!Williams, R.M and Tuckey,P.A. (1992)!
     Regge Calculus : A bibliography and a brief review!
     Class.Q.Grav. Vol.9(1992) pp.1409-1422.\par

\R 3!Gentle, A.P. and Miller, W.A. (1997)!
     A fully (3+1)-d evolution of the Kasner cosmology using the Sorkin scheme!
     Submitted Class.Q.Grav.\par

\R 4!Miller,M.A. (1995)!
     Regge Calculus as a fourth order method in Numerical Relativity.!
     Class.Q.Grav. Vol.12(1995) pp.3037-3051.\par

\R 5!Richtmyer, R.D. and Morton, K.W. (1967)!
     Difference methods for initial value problems. 2nd edition.!
     Wiley-Interscience, New York, 1967.\par

\R 6!Brewin, L.C. (1997)!
     A Finite Element formulation of Simplicial General Relativity.!
     Preprint, Mathematics Department, Monash University.\hfil\break
     Also available at {\tt http://newton.maths.monash.edu.au:8000/preprints/%
     felm-gr.ps.gz}\par

\R 7!Brewin, L.C. (1993)!
     Particle paths in a Schwarzschild spacetime via the Regge Calculus.!
     Class.Q.Grav. Vol.10(1993) pp.1803-1823.\par

\R 8!Lewis, S.M. (1972)!
     Two cosmological solutions of Regge Calculus.!
     Phys.Rev.D. Vol.25(1982) pp.306-312.\par

\R 9!Cheuk-yin Wong. (1971)!
     Application of Regge Calculus to the Schwarzschild and Reissner--Nordstrom
     geometries at the moment of time symmetry.!
     J.Math.Phys. Vol.12(1971) pp.70-78.\par

\R10!Barrett, J.W. (1988)!
     A convergence result for linearised Regge Calculus.!
     Class.Q.Grav. Vol.5(1988) pp.1187-1192.\par

\R11!Hardy, G.H. (1958)!
     A course of pure mathematics.!
     Cambridge University Press (1958).\par

\egroup
\vfill\eject
%-------------------------------------------------------------------------------
\twelvepointsgl
\nopagenumbers
\hsize=19.65cm
\hoffset=-1.5cm
%
% two types of horizontal lines
%
\def\maindivider{\leaders\hrule height 1.5pt depth 0pt\hfil}
\def\medmdivider{\leaders\hrule height 1.0pt depth 0pt\hfil}
\def\thindivider{\leaders\hrule height 0.5pt depth 0pt\hfil}
\def\Stroke{\vrule width 1.0pt}
\def\MyStroke{\omit\Stroke}
%
% a bit of vertical space
%

%
% table = \hbox { left hand edge, \vbox, right hand edge }
% the \vbox contains the body of the table
%
% \tabskip is the inter-column hglue. setting it in one column
% will only have effect in subsequent columns
%
% \offinterlineskip is used in conjunction with the \vrule entry
% in \halign to force each line in the table to be of a specific
% height and depth
%
% NOTE : At least one line in the \halign must have an entry for ALL
%        of the fields declared in the template (including the final
%        vrule entry, I do this by ending some lines with &\cr not \cr).
% NOTE : The integer parameter in the multispan equals the number of
%        columns in the table plus two.
%
%
% -------- Model 1 ---------- aka : Simple
%
\newbox\boxa
\newbox\boxb
\newbox\boxc
\newbox\boxd
\offinterlineskip
\setbox\boxa=\hbox{\hfil\vrule width 1.5pt%
\vtop{\twelvepoint\offinterlineskip\tabskip 0.0cm\halign{%
\vrule height 16pt depth 7pt width 0pt#\tabskip=0.40cm&%
\hfil{#}\hfil&#\vrule width 0.5pt&%
\hfil{#}\hfil&#\vrule width 0.5pt&%
\hfil{#}\hfil&#\vrule width 0.5pt&%
\hfil{#}\hfil&#\vrule width 0.5pt&%
\hfil{#}\hfil&%
\vrule height 16pt depth 7pt width 0pt#\tabskip=0.0cm\cr
%--- end of template
\multispan{11}\maindivider\cr
&\multispan{9}\hfil \bf Coordinates for Model 1\hfil&\cr
\multispan{11}\maindivider\cr
&\bf Vertex&\MyStroke&$(x^1)$&&$(x^2)$&&$(x^3)$&&$(x^4)$&\cr
\multispan{11}\medmdivider\cr
&1&\MyStroke&0.500&&0.500&&0.500&&0.500&\cr
\multispan{11}\thindivider\cr
&2&\MyStroke&0.500&&1.000&&0.500&&0.500&\cr
\multispan{11}\thindivider\cr
&3&\MyStroke&0.500&&0.500&&1.000&&0.500&\cr
\multispan{11}\thindivider\cr
&4&\MyStroke&0.500&&0.500&&0.500&&1.000&\cr
\multispan{11}\thindivider\cr
&5&\MyStroke&0.500&&0.625&&0.625&&0.625&\cr
\multispan{11}\thindivider\cr
&$5^\up$&\MyStroke&1.000&&0.625&&0.625&&0.625&\cr
\multispan{11}\thindivider\cr
&$5^\dn$&\MyStroke&0.000&&0.625&&0.625&&0.625&\cr
\multispan{11}\thindivider\cr
&$V_{\star}$&\MyStroke&0.500&&0.625&&0.625&&0.625&\cr
\multispan{11}\maindivider\cr}}\vrule width 1.5pt\hfil}
%
% -------- Model 2 ---------- aka : hills
%
\setbox\boxb=\hbox{\hfil\vrule width 1.5pt%
\vtop{\twelvepoint\offinterlineskip\tabskip 0.0cm\halign{%
\vrule height 16pt depth 7pt width 0pt#\tabskip=0.40cm&%
\hfil{#}\hfil&#\vrule width 0.5pt&%
\hfil{#}\hfil&#\vrule width 0.5pt&%
\hfil{#}\hfil&#\vrule width 0.5pt&%
\hfil{#}\hfil&#\vrule width 0.5pt&%
\hfil{#}\hfil&%
\vrule height 16pt depth 7pt width 0pt#\tabskip=0.0cm\cr
%--- end of template
\multispan{11}\maindivider\cr
&\multispan{9}\hfil \bf Coordinates for Model 2\hfil&\cr
\multispan{11}\maindivider\cr
&\bf Vertex&\MyStroke&$(x^1)$&&$(x^2)$&&$(x^3)$&&$(x^4)$&\cr
\multispan{11}\medmdivider\cr
&1&\MyStroke&0.500&&0.500&&0.500&&0.500&\cr
\multispan{11}\thindivider\cr
&2&\MyStroke&0.500&&0.500&&0.500&&1.000&\cr
\multispan{11}\thindivider\cr
&3&\MyStroke&0.500&&0.500&&0.500&&0.000&\cr
\multispan{11}\thindivider\cr
&4&\MyStroke&0.500&&1.000&&0.500&&0.500&\cr
\multispan{11}\thindivider\cr
&5&\MyStroke&0.500&&0.500&&1.000&&0.500&\cr
\multispan{11}\thindivider\cr
&6&\MyStroke&0.500&&0.000&&0.500&&0.500&\cr
\multispan{11}\thindivider\cr
&7&\MyStroke&0.500&&0.500&&0.000&&0.500&\cr
\multispan{11}\thindivider\cr
&$1^\up$&\MyStroke&1.000&&0.500&&0.500&&0.500&\cr
\multispan{11}\thindivider\cr
&$1^\dn$&\MyStroke&0.000&&0.500&&0.500&&0.500&\cr
\multispan{11}\thindivider\cr
&$V_{\star}$&\MyStroke&0.500&&0.500&&0.500&&0.500&\cr
\multispan{11}\maindivider\cr}}\vrule width 1.5pt\hfil}
%
% -------- Model 3 ---------- aka : hcube
%
\setbox\boxc=\hbox{\hfil\vrule width 1.5pt%
\vtop{\twelvepoint\offinterlineskip\tabskip 0.0cm\halign{%
\vrule height 16pt depth 7pt width 0pt#\tabskip=0.40cm&%
\hfil{#}\hfil&#\vrule width 0.5pt&%
\hfil{#}\hfil&#\vrule width 0.5pt&%
\hfil{#}\hfil&#\vrule width 0.5pt&%
\hfil{#}\hfil&#\vrule width 0.5pt&%
\hfil{#}\hfil&%
\vrule height 16pt depth 7pt width 0pt#\tabskip=0.0cm\cr
%--- end of template
\multispan{11}\maindivider\cr
&\multispan{9}\hfil \bf Coordinates for Model 3\hfil&\cr
\multispan{11}\maindivider\cr
&\bf Vertex&\MyStroke&$(x^1)$&&$(x^2)$&&$(x^3)$&&$(x^4)$&\cr
\multispan{11}\medmdivider\cr
&0--255&\MyStroke&\multispan{7}\hfil See text\hfil&\cr
\multispan{11}\thindivider\cr
&$V_{\star}$&\MyStroke&0.250&&1.000&&1.000&&1.000&\cr
\multispan{11}\maindivider\cr}}\vrule width 1.5pt\hfil}
\setbox\boxd=\hbox{\hbox to 0.45\hsize{\hfil\hsize=0.45\hsize\twelvepoint
\advance\baselineskip+1.7pt%
\vtop{\vfil{\bf Table \tbldef{VerticesTable}.}\ Coordinates of the vertices of
the various models.
\vfil}\hfil}}
\hbox{%
\vtop{\box\boxa\vskip28pt\box\boxd}\hskip8pt%
\vtop{\box\boxb\vskip 8pt\box\boxc}}

\vfill\eject

\input rotate
\input BoxedEPS
\HideDisplacementBoxes
\SetRokickiEPSFSpecial
\twelvepointsgl
\nopagenumbers
\def\eps{\epsilon}
\def\Oeps(#1){{\cal O}(\eps^{#1})}

\def\OZero{{\cal O}(1)}

\def\MyBig{\seventeenpointsgl}

\advance\baselineskip 2pt
\advance\vsize 1.0cm\relax
%
%-----------------------------------------------------------------------------
\def\stomp#1{\setbox0=\hbox{#1}\dp0=0pt\ht0=0pt\box0}
%
% This version of the \at routine ensures that its boxe's
% have no depth, width or height.
%
\def\at(#1,#2)#3{\vbox to 0pt{\kern#2cm%
                 \hbox to 0pt{\kern#1cm\stomp{#3}\hss}\vss}\nointerlineskip}
%-----------------------------------------------------------------------------
%
\parindent=0pt
\newbox\boxa
\newbox\ylabel
\newbox\xlabel
\newbox\heada
\newbox\headb
\newbox\headc
\newbox\ModelOne
\newbox\ModelTwo
\newbox\ModelThree
\setbox\ModelOne=\hbox to 8cm{\hsize=8cm\vtop{%
{\bf Figure \figdef{ModelOne}.} The first model, a simple collection of 4 tetrahedra. The full
4-dimensional simplicial space is obtained by displacing the central
vertex forward and backward in time.}}
\setbox\ModelTwo=\hbox to 8cm{\hsize=8cm\vtop{%
{\bf Figure \figdef{ModelTwo}.} The second model. In this model the central vertex is also
dragged forward and backward in time.}}
\setbox\ModelThree=\hbox to 6cm{\hsize=6cm\vtop{\raggedright%
{\bf Figure \figdef{ModelThree}.} An example of a three dimensional hypercube. The heavy dotted
lines are part of the template for this hypercube. The light dotted lines
arise form templates of other hypercubes.}}
\setbox\boxa=%
\hbox to 15.0cm{% Specifies the width of the box
\vtop to 18.5cm{% Specifies the depth of the box, the box has zero height.
\at(  0.0, 24.0){\bBoxedEPSF{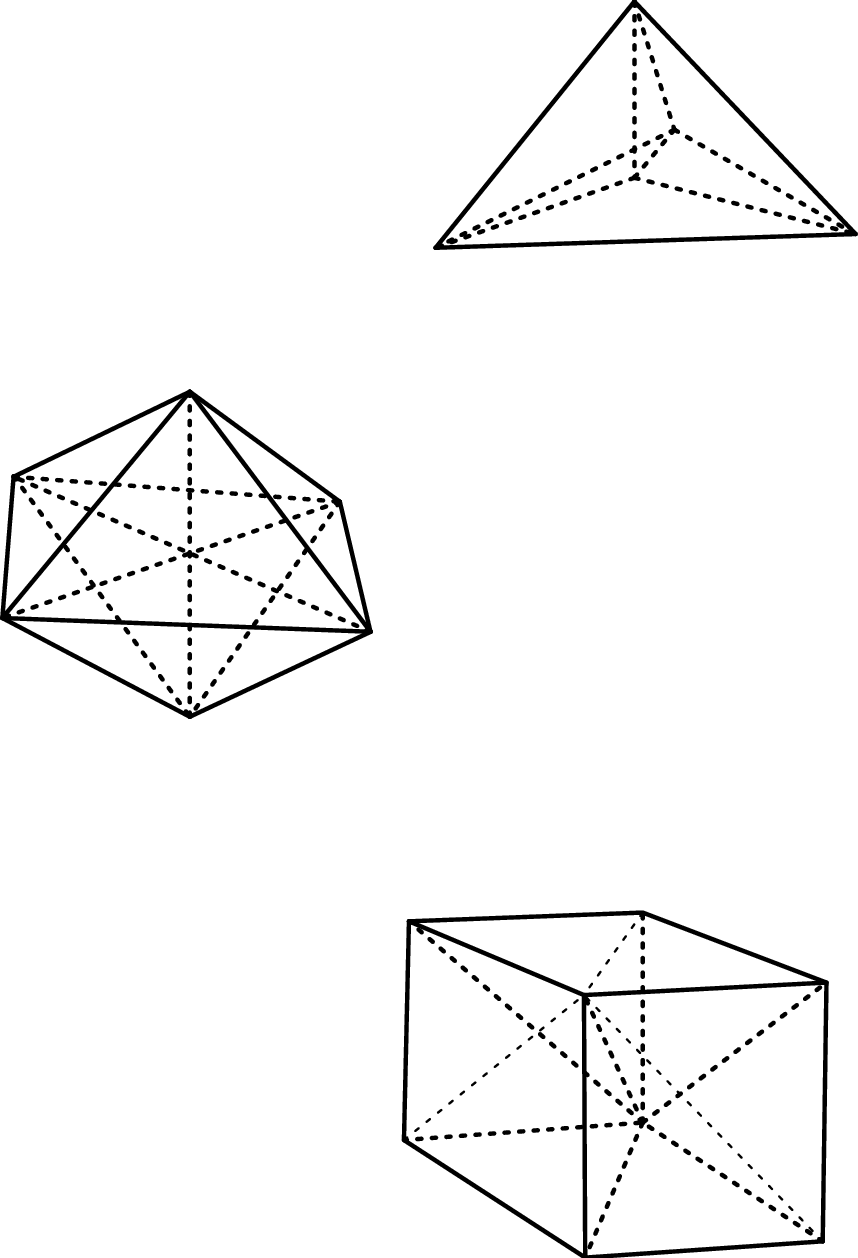}}
\at(  0.0,  3.5){\box\ModelOne}
\at(  8.0, 12.0){\box\ModelTwo}
\at(  0.0, 19.0){\box\ModelThree}
\vfill}\hfill}
%
% Now place the boxes on the page.
%
\centerline{\box\boxa}
\vfill\eject
%-----------------------------------------------------------------------------
\setbox\ylabel=\hbox{\MyBig $d{\rm Log}\eta\strut_2/d{\rm Log}\epsilon$}
\setbox\xlabel=\hbox{\MyBig $-{\rm Log}\,\epsilon$}
\setbox\ylabel=\hbox{\rotl\ylabel}
\setbox\heada=\hbox{\MyBig Effective Residual Errors, Model 1}
\setbox\headb=\hbox{\MyBig Rms}
\setbox\headc=\hbox{\MyBig Max}
\setbox\boxa=%
\hbox to 15.0cm{% Specifies the width of the box
\vtop to 18.5cm{% Specifies the depth of the box, the box has zero height.
\at( 0.0,15.5){\bBoxedEPSF{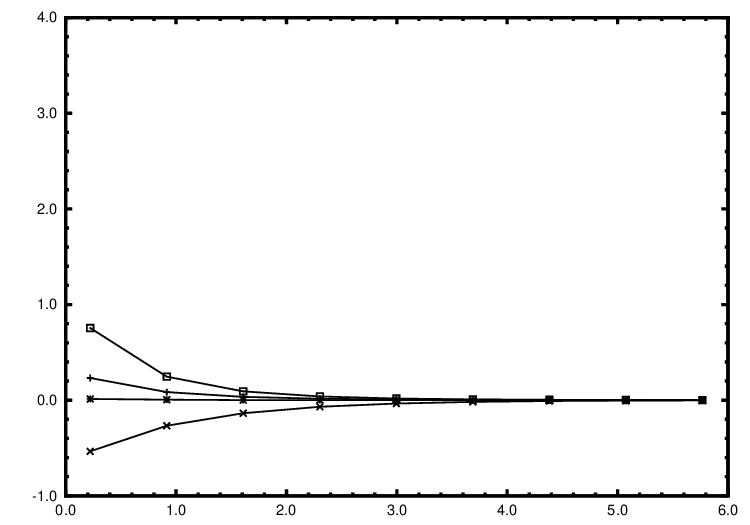 scaled 1150}}
\at( 6.8,16.5){\box\xlabel}%6.8,16.5
\at(-0.75,11.5){\box\ylabel}
\at( 3.0, 2.5){\box\heada}
\vfill}\hfill}
%
% Now place the boxes on the page.
%
\centerline{\box\boxa}
\vskip 1cm
\centerline{\vtop{\hsize=12cm{\bf Figure \figdef{FigD}.}
This graph displays $d\log\eta\strut_2/d\log\eps$ as a function of $-\log\eps$
for the original Regge equations for Model 1. The squares correspond to 
the Schwarzschild spacetime, triangles to the Kasner solution, diamonds to
the plane wave, stars to the symmetric wave. Note that for every metric the
effective residual error varies as $\OZero$.
The derivatives were estimated using
$d\eta/d\epsilon = (\eta_{i+1}-\eta_i)/(\epsilon_{i+1}-\epsilon_i)$. 
}}\vfill\eject
%-----------------------------------------------------------------------------
\setbox\ylabel=\hbox{\MyBig $d{\rm Log}\eta\strut_2/d{\rm Log}\epsilon$}
\setbox\xlabel=\hbox{\MyBig $-{\rm Log}\,\epsilon$}
\setbox\ylabel=\hbox{\rotl\ylabel}
\setbox\heada=\hbox{\MyBig Effective Residual Errors, Model 2}
\setbox\headb=\hbox{\MyBig Rms}
\setbox\headc=\hbox{\MyBig Max}
\setbox\boxa=%
\hbox to 15.0cm{% Specifies the width of the box
\vtop to 18.5cm{% Specifies the depth of the box, the box has zero height.
\at( 0.0,15.5){\bBoxedEPSF{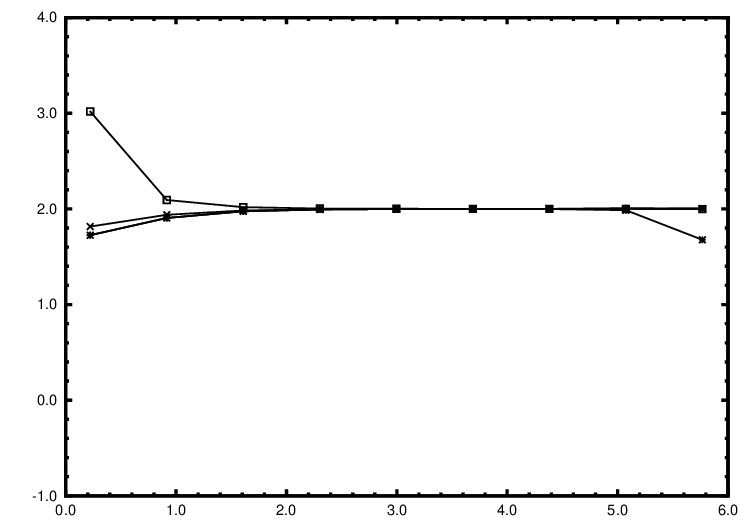 scaled 1150}}
\at( 6.8,16.5){\box\xlabel}
\at(-0.75,11.5){\box\ylabel}
\at( 3.0, 2.5){\box\heada}
\vfill}\hfill}
%
% Now place the boxes on the page.
%
\centerline{\box\boxa}
\vskip 1cm
\centerline{\vtop{\hsize=12cm{\bf Figure \figdef{FigE}.}
Results for the original Regge equations for Model 2. 
For the Kasner and both plane wave metrics the effective residual error varies
as $\Oeps(2)$ while for the Schwarzschild metric we only
have $\OZero$.
The dip in the curves
near $5<-{\rm Log}\epsilon<6$ indicates that machine precision has been reached.
This behaviour can also be seen in many of the other plots.
}}\vfill\eject
%-----------------------------------------------------------------------------
\setbox\ylabel=\hbox{\MyBig $d{\rm Log}\eta\strut_2/d{\rm Log}\epsilon$}
\setbox\xlabel=\hbox{\MyBig $-{\rm Log}\,\epsilon$}
\setbox\ylabel=\hbox{\rotl\ylabel}
\setbox\heada=\hbox{\MyBig Effective Residual Errors, Model 3}
\setbox\headb=\hbox{\MyBig Rms}
\setbox\headc=\hbox{\MyBig Max}
\setbox\boxa=%
\hbox to 15.0cm{% Specifies the width of the box
\vtop to 18.5cm{% Specifies the depth of the box, the box has zero height.
\at( 0.0,15.5){\bBoxedEPSF{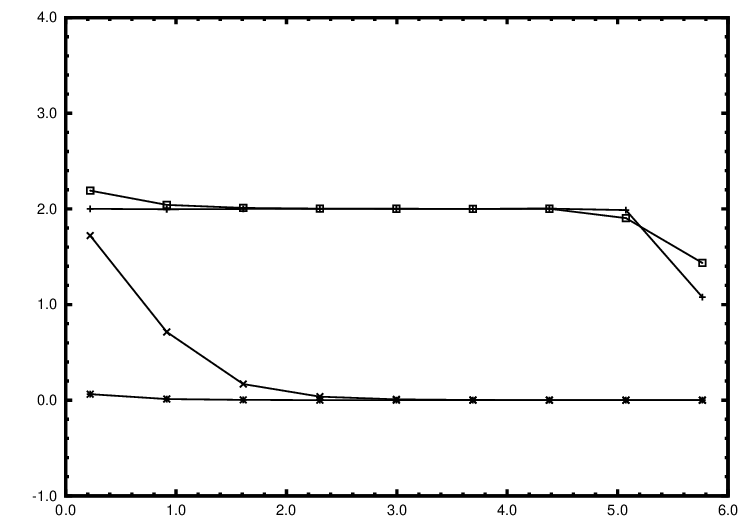 scaled 1150}}
\at( 6.8,16.5){\box\xlabel}
\at(-0.75,11.5){\box\ylabel}
\at( 3.0, 2.5){\box\heada}
\vfill}\hfill}
%
% Now place the boxes on the page.
%
\centerline{\box\boxa}
\vskip 1cm
\centerline{\vtop{\hsize=12cm{\bf Figure \figdef{FigF}.}
Results for the original Regge equations for Model 3. This displays an
$\Oeps(2)$ effective residual error for only the Kasner and plane wave
metrics.
}}\vfill\eject
%-----------------------------------------------------------------------------
%-----------------------------------------------------------------------------
\setbox\ylabel=\hbox{\MyBig $d{\rm Log}\eta\strut_2/d{\rm Log}\epsilon$}
\setbox\xlabel=\hbox{\MyBig $-{\rm Log}\,\epsilon$}
\setbox\ylabel=\hbox{\rotl\ylabel}
\setbox\heada=\hbox{\MyBig Effective Residual Errors, Model 1}
\setbox\headb=\hbox{\MyBig Rms}
\setbox\headc=\hbox{\MyBig Max}
\setbox\boxa=%
\hbox to 15.0cm{% Specifies the width of the box
\vtop to 18.5cm{% Specifies the depth of the box, the box has zero height.
\at( 0.0,15.5){\bBoxedEPSF{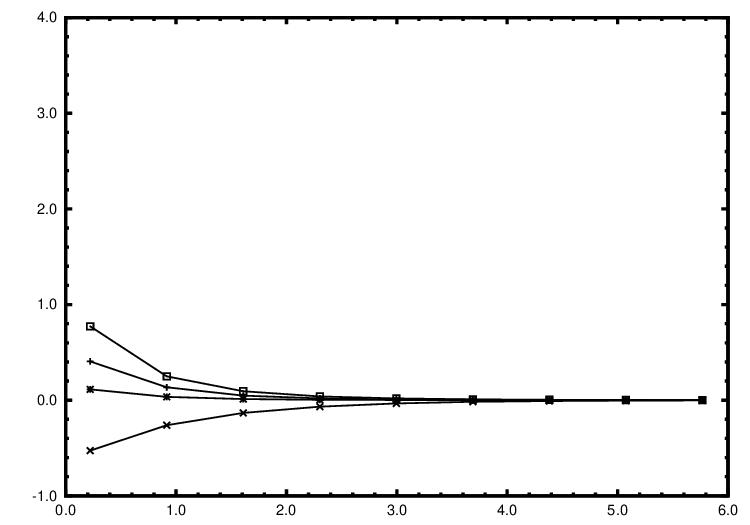 scaled 1150}}
\at( 6.8,16.5){\box\xlabel}
\at(-0.75,11.5){\box\ylabel}
\at( 3.0, 2.5){\box\heada}
\vfill}\hfill}
%
% Now place the boxes on the page.
%
\centerline{\box\boxa}
\vskip 1cm
\centerline{\vtop{\hsize=12cm{\bf Figure \figdef{FigG}.}
Results for the Brewin's equations for Model 1.
}}\vfill\eject
%-----------------------------------------------------------------------------
\setbox\ylabel=\hbox{\MyBig $d{\rm Log}\eta\strut_2/d{\rm Log}\epsilon$}
\setbox\xlabel=\hbox{\MyBig $-{\rm Log}\,\epsilon$}
\setbox\ylabel=\hbox{\rotl\ylabel}
\setbox\heada=\hbox{\MyBig Effective Residual Errors, Model 2}
\setbox\headb=\hbox{\MyBig Rms}
\setbox\headc=\hbox{\MyBig Max}
\setbox\boxa=%
\hbox to 15.0cm{% Specifies the width of the box
\vtop to 18.5cm{% Specifies the depth of the box, the box has zero height.
\at( 0.0,15.5){\bBoxedEPSF{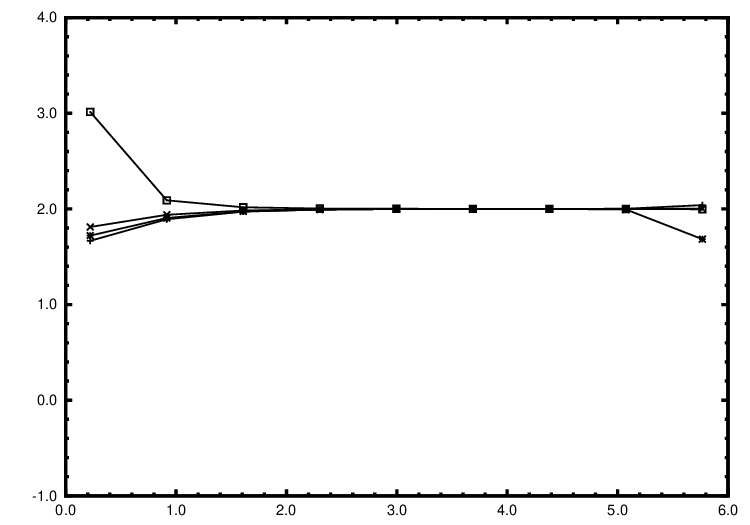 scaled 1150}}
\at( 6.8,16.5){\box\xlabel}
\at(-0.75,11.5){\box\ylabel}
\at( 3.0, 2.5){\box\heada}
\vfill}\hfill}
%
% Now place the boxes on the page.
%
\centerline{\box\boxa}
\vskip 1cm
\centerline{\vtop{\hsize=12cm{\bf Figure \figdef{FigH}.}
Results for the Brewin's equations for Model 2.
}}\vfill\eject
%-----------------------------------------------------------------------------
\setbox\ylabel=\hbox{\MyBig $d{\rm Log}\eta\strut_2/d{\rm Log}\epsilon$}
\setbox\xlabel=\hbox{\MyBig $-{\rm Log}\,\epsilon$}
\setbox\ylabel=\hbox{\rotl\ylabel}
\setbox\heada=\hbox{\MyBig Effective Residual Errors, Model 3}
\setbox\headb=\hbox{\MyBig Rms}
\setbox\headc=\hbox{\MyBig Max}
\setbox\boxa=%
\hbox to 15.0cm{% Specifies the width of the box
\vtop to 18.5cm{% Specifies the depth of the box, the box has zero height.
\at( 0.0,15.5){\bBoxedEPSF{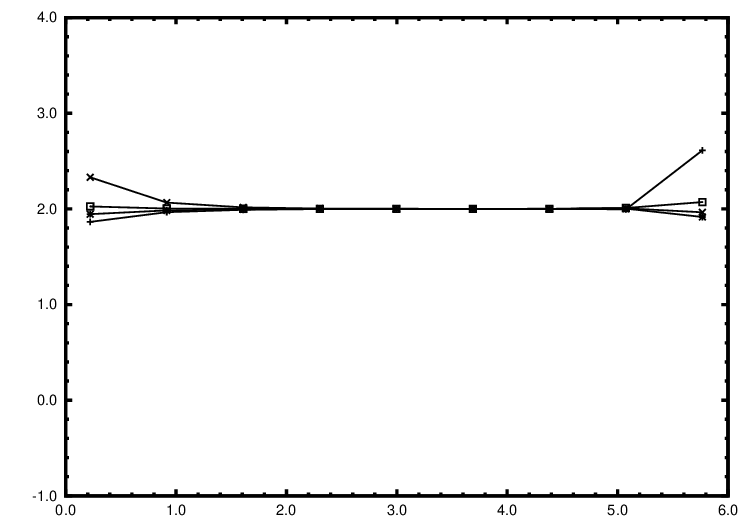 scaled 1150}}
\at( 6.8,16.5){\box\xlabel}
\at(-0.75,11.5){\box\ylabel}
\at( 3.0, 2.5){\box\heada}
\vfill}\hfill}
%
% Now place the boxes on the page.
%
\centerline{\box\boxa}
\vskip 1cm
\centerline{\vtop{\hsize=12cm{\bf Figure \figdef{FigI}.}
Results for the Brewin's equations for Model 3.
}}\vfill\eject
%-----------------------------------------------------------------------------
%-----------------------------------------------------------------------------
\setbox\ylabel=\hbox{\MyBig $d{\rm Log}\eta\strut_2/d{\rm Log}\epsilon$}
\setbox\xlabel=\hbox{\MyBig $-{\rm Log}\,\epsilon$}
\setbox\ylabel=\hbox{\rotl\ylabel}
\setbox\heada=\hbox{\MyBig Effective Residual Errors, Model 1}
\setbox\headb=\hbox{\MyBig Rms}
\setbox\headc=\hbox{\MyBig Max}
\setbox\boxa=%
\hbox to 15.0cm{% Specifies the width of the box
\vtop to 18.5cm{% Specifies the depth of the box, the box has zero height.
\at( 0.0,15.5){\bBoxedEPSF{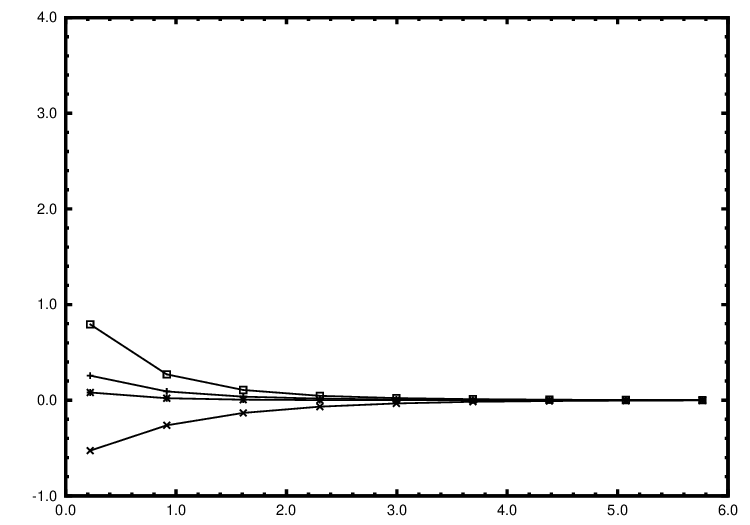 scaled 1150}}
\at( 6.8,16.5){\box\xlabel}
\at(-0.75,11.5){\box\ylabel}
\at( 3.0, 2.5){\box\heada}
\vfill}\hfill}
%
% Now place the boxes on the page.
%
\centerline{\box\boxa}
\vskip 1cm
\centerline{\vtop{\hsize=12cm{\bf Figure \figdef{FigJ}.}
Results for Miller's equations (modified as per section (\secrfr{MillersEqtns}))
for Model 1.
}}\vfill\eject
%-----------------------------------------------------------------------------
\setbox\ylabel=\hbox{\MyBig $d{\rm Log}\eta\strut_2/d{\rm Log}\epsilon$}
\setbox\xlabel=\hbox{\MyBig $-{\rm Log}\,\epsilon$}
\setbox\ylabel=\hbox{\rotl\ylabel}
\setbox\heada=\hbox{\MyBig Effective Residual Errors, Model 2}
\setbox\headb=\hbox{\MyBig Rms}
\setbox\headc=\hbox{\MyBig Max}
\setbox\boxa=%
\hbox to 15.0cm{% Specifies the width of the box
\vtop to 18.5cm{% Specifies the depth of the box, the box has zero height.
\at( 0.0,15.5){\bBoxedEPSF{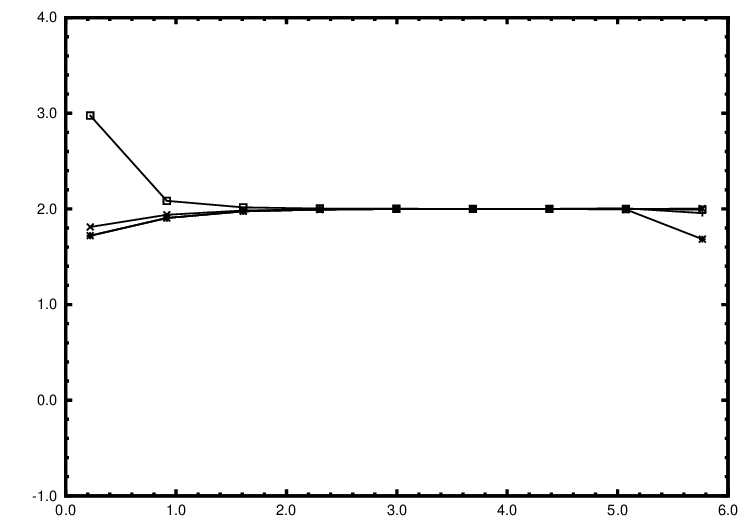 scaled 1150}}
\at( 6.8,16.5){\box\xlabel}
\at(-0.75,11.5){\box\ylabel}
\at( 3.0, 2.5){\box\heada}
\vfill}\hfill}
%
% Now place the boxes on the page.
%
\centerline{\box\boxa}
\vskip 1cm
\centerline{\vtop{\hsize=12cm{\bf Figure \figdef{FigK}.}
Results for Miller's equations (modified as per section (\secrfr{MillersEqtns}))
for Model 2.
}}\vfill\eject
%-----------------------------------------------------------------------------
\setbox\ylabel=\hbox{\MyBig $d{\rm Log}\eta\strut_2/d{\rm Log}\epsilon$}
\setbox\xlabel=\hbox{\MyBig $-{\rm Log}\,\epsilon$}
\setbox\ylabel=\hbox{\rotl\ylabel}
\setbox\heada=\hbox{\MyBig Effective Residual Errors, Model 3}
\setbox\headb=\hbox{\MyBig Rms}
\setbox\headc=\hbox{\MyBig Max}
\setbox\boxa=%
\hbox to 15.0cm{% Specifies the width of the box
\vtop to 18.5cm{% Specifies the depth of the box, the box has zero height.
\at( 0.0,15.5){\bBoxedEPSF{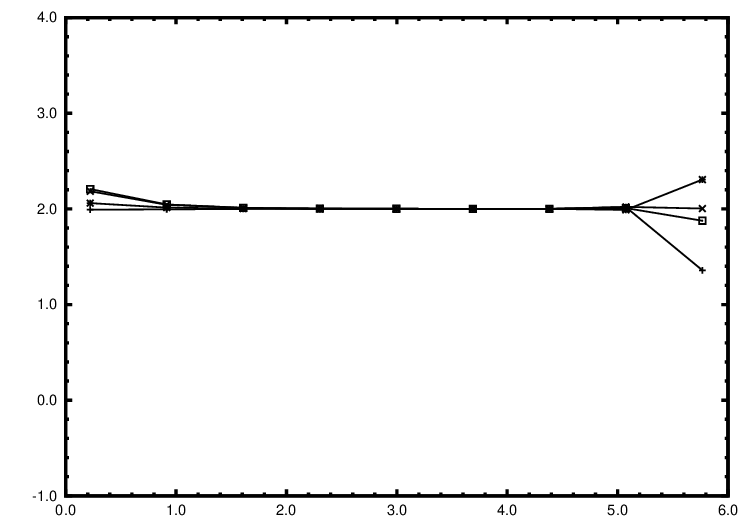 scaled 1150}}
\at( 6.8,16.5){\box\xlabel}
\at(-0.75,11.5){\box\ylabel}
\at( 3.0, 2.5){\box\heada}
\vfill}\hfill}
%
% Now place the boxes on the page.
%
\centerline{\box\boxa}
\vskip 1cm
\centerline{\vtop{\hsize=12cm{\bf Figure \figdef{FigL}.}
Results for Miller's equations for Model 3.
}}\vfill\eject
%-----------------------------------------------------------------------------

\bye

%% file: preprint.tex
\input 10ptsgl
\input 12ptdbl
\input 17ptdbl
\twelvepointsgl
\newbox\pgno
\def\pagenumbers{%
\def\folio{\ifnum\pageno>1%
\setbox\pgno=\vbox{\kern+0.5cm\hbox{{\tt-}\number\pageno{\tt-}}}%
\dp\pgno=0pt\ht\pgno=0pt\box\pgno\fi}
\footline={\twelvepoint\hss\folio\hss}%
\headline={\hfil}}
\def\nopagenumbers{%
\def\folio{\null}
\headline={\hfil}
\footline={\hfil}}
\pagenumbers
\vsize=23cm%  for A4 paper (ie. Oz,UK,etc.)
\baselineskip 16pt%       % was 15pt
\lineskip      2pt%       % was 2pt
\lineskiplimit 1pt%
\parindent     0pt%
\parskip       5pt plus 4pt minus 1pt% was 5pt plus 4pt
\def\begnarrow{\par\advance\leftskip20pt}
\def\endnarrow{\par\advance\leftskip-20pt}
\def\smlskip{\vskip  3pt plus1pt minus1pt}
\def\medskip{\vskip  6pt plus2pt minus2pt}
\def\bigskip{\vskip 12pt plus4pt minus4pt}
%

%
% --- Define macros for title, address, abstract and sections -----------------
%
\def\TheSecNum{0}%
\def\title#1{{\seventeenpoint\advance\baselineskip5pt\bf%
               \halign{\hbox to \hsize{\hfil##\hfil}\cr#1}}}
\def\address#1{\halign{\hbox to \hsize{\hfil##\hfil}\cr#1}}
\def\beginabstract{\par\bgroup\tenpoint\vskip1cm%
\centerline{\twelvepoint\bf Abstract}%
\advance\leftskip0cm\advance\rightskip0cm\smlskip}%
\def\endabstract{\par\egroup}
\newcount\secnum
\newcount\subsecnum
\newcount\subsubsecnum
\outer\def\beginsection#1{%
\vskip+4\baselineskip\penalty-500%
\vskip-4\baselineskip%
\vskip 20pt plus 6pt minus 3pt%
\advance\secnum +1\subsecnum=0\subsubsecnum=0\equationnum=0%
{\seventeenpoint\bf\the\secnum.\ #1}%
\def\TheSecNum{\the\secnum}%
\vskip 5pt}%
\outer\def\beginsubsection#1{%
\vskip+4\baselineskip\penalty-500%
\vskip-4\baselineskip%
\vskip 10pt plus 3pt minus 2pt%
\advance\subsecnum +1\subsubsecnum=0\equationnum=0%
{\twelvepoint\bf\the\secnum.\the\subsecnum.\ #1}%
\def\TheSecNum{\the\secnum.\the\subsecnum}%
\vskip 5pt}%
\outer\def\beginsubsubsection#1{%
\vskip+4\baselineskip\penalty-500%
\vskip-4\baselineskip%
\vskip 10pt plus 3pt minus 2pt%
\advance\subsubsecnum +1%
{\twelvepoint\bf\the\secnum.\the\subsecnum.\the\subsubsecnum.\ #1}%
\def\TheSecNum{\the\secnum.\the\subsecnum}%
\vskip 5pt}%
%
% --- Silently read a file ----------------------------------------------------
%
\newread\tmpfile
%

%
% --- Aux and Bib files -------------------------------------------------------
%
\newread\auxfile
\openin\auxfile=./\jobname.aux
\ifeof\auxfile\def\readaux{\relax}\else
              \def\readaux{\input ./\jobname.aux\relax}\fi
\closein\auxfile
\readaux\def\readaux{\relax}
\newwrite\auxfile
\immediate\openout\auxfile=./\jobname.aux
\newwrite\bibfile
\immediate\openout\bibfile=./\jobname.bib
\def\readbib{\input ./\jobname.bib\relax}
\def\ReadBib{\immediate\closeout\bibfile\readbib\def\ReadBib{\relax}}
\def\WriteBib{\immediate\closeout\bibfile\def\WriteBib{\relax}}
%
% --- Define macros for equation numbers --------------------------------------
%
\newcount \equationnum \equationnum = 0
\def\numberequation #1{\global\advance\equationnum by 1%
    \expandafter\xdef\csname eqn#1\endcsname{\TheSecNum.\the\equationnum}%
\immediate\write\auxfile{%
    \def\expandafter\noexpand\csname eqn#1\endcsname{%
        \TheSecNum.\the\equationnum}}}

\def\labelequation #1{\expandafter\xdef\csname eqn#1\endcsname{#1}}

\def\eqnlabel#1{\numberequation{#1}}% --- Use this for numbered equations

\def\eqndef#1{\ignorespaces\eqnlabel{#1}(\csname eqn#1\endcsname)}
\def\eqnrfr#1{\csname eqn#1\endcsname}% 5 Dec 2006, was wrapped in ( )
\def\eqnnxt{\global\advance\equationnum by 1%
           {(\TheSecNum.\the\equationnum)}}
%
% --- Define macros for figure numbers ----------------------------------------
%
\newcount \figurenum \figurenum = 0
\def\numberfigure #1{\global\advance\figurenum by 1%
    \expandafter\xdef\csname fig#1\endcsname{\the\figurenum}%
\immediate\write\auxfile{%
    \def\expandafter\noexpand\csname fig#1\endcsname{%
        \the\figurenum}}}

\def\labelfigure #1{\expandafter\xdef\csname fig#1\endcsname{#1}}

\def\figlabel#1{\numberfigure{#1}}% --- Use this for numbered figures

\def\figdef#1{\ignorespaces\figlabel{#1}\csname fig#1\endcsname}
\def\figrfr#1{\csname fig#1\endcsname}
%
% --- Define macros for table numbers -----------------------------------------
%
\newcount \tablenum \tablenum = 0
\def\numbertable #1{\global\advance\tablenum by 1%
    \expandafter\xdef\csname tbl#1\endcsname{\the\tablenum}%
\immediate\write\auxfile{%
    \def\expandafter\noexpand\csname tbl#1\endcsname{%
        \the\tablenum}}}

\def\labeltable #1{\expandafter\xdef\csname tbl#1\endcsname{#1}}

\def\tbllabel#1{\numbertable{#1}}% --- Use this for numbered tables

\def\tbldef#1{\ignorespaces\tbllabel{#1}\csname tbl#1\endcsname}
\def\tblrfr#1{\csname tbl#1\endcsname}
%
% --- Define macros for section numbers ---------------------------------------
%
\def\numbersection #1{%
    \expandafter\xdef\csname sec#1\endcsname{\TheSecNum}%
\immediate\write\auxfile{%
    \def\expandafter\noexpand\csname sec#1\endcsname{%
        \TheSecNum}}}

\def\labelsection #1{\expandafter\xdef\csname sec#1\endcsname{#1}}

\def\seclabel#1{\numbersection{#1}}% --- Use this for numbered sections

\def\secdef#1{\ignorespaces\seclabel{#1}}%
\def\secrfr#1{\csname sec#1\endcsname}%
%
% --- Define macros for citations etc -----------------------------------------
%
\newcount \citenum \citenum = 0
\def\citedef #1#2{%
    \expandafter\xdef\csname cite#1\endcsname{\noexpand\TriggerCite{#1}}%
    \expandafter\xdef\csname cite#1Long\endcsname%
    {\noexpand\REF\expandafter\noexpand\csname cite#1\endcsname!#2}%
\immediate\write\auxfile{%
    \def\expandafter\noexpand\csname cite#1\endcsname{\the\citenum}}%
\immediate\write\auxfile{%
    \def\expandafter\noexpand\csname cite#1Long\endcsname%
    {\noexpand\REF\expandafter\noexpand\csname cite#1\endcsname!#2}}}

\def\cite#1{\CiteLeft\csname cite#1\CiteNameTail\endcsname\CiteRight}

\def\TriggerCite#1{%
\global\advance\citenum by 1%
\expandafter\xdef\csname cite#1\endcsname{\the\citenum}%
\csname cite#1\endcsname%
\immediate\write\auxfile{%
    \def\expandafter\noexpand\csname cite#1\endcsname{\the\citenum}}%
\immediate\write\bibfile{%
    \noexpand\cite{#1}}}
%
% -----------------------------------------------------------------------------
%
\outer\def\testbreak{%
\vskip0pt plus+2\baselineskip\penalty-250%
\vskip0pt plus-2\baselineskip}
\def\RR #1!#2!#3\par{\vskip 0pt plus .1\vsize\penalty -250%
                     \vskip 0pt plus-.1\vsize%
                     \rm\item{\bf [#1]}%              % Ref. number
                     \rm\ignorespaces #2\par\item{}%  % Author
                        \ignorespaces #3\par\par\rm%  % Reference
                     \vskip 5pt}
\def\R #1!#2!#3!#4\par{\vskip 0pt plus .1\vsize\penalty -250%
                       \vskip 0pt plus-.1\vsize%
                       \rm\item{\bf [#1]}%              % Ref. number
                       \rm\ignorespaces #2\par\item{}%  % Author
                       \sl\ignorespaces #3\par\item{}%  % Title
                       \rm\ignorespaces #4\par\par\rm%  % Reference
                       \vskip 5pt}
\def\REF #1!#2!#3!#4!{\vskip 0pt plus .1\vsize\penalty -250%
                      \vskip 0pt plus-.1\vsize%
                      \rm\item{\bf [#1]}%                % Ref. number
                      \rm\ignorespaces #2\par%           % Author
                      \sl\ignorespaces #3\par%           % Title
                      \rm\ignorespaces #4\par\rm%        % Reference
                      \vskip 7pt}

%% file: 12ptdbl.tex
%
% Double space macro for 12 point font.
%
    \gdef\twelvepointdbl{%
         \twelvepoint%
         \abovedisplayskip 18pt plus 4pt minus 6pt%
         \belowdisplayskip 18pt plus 4pt minus 6pt%
         \abovedisplayshortskip  0pt plus 4pt%
         \belowdisplayshortskip 10pt plus 4pt minus 4pt%
         \baselineskip 22pt%
         \lineskip     11pt%
         \lineskiplimit 3pt%
         \parindent     0pt%
         \parskip       0pt plus 2pt}%
%
%
% Single space macro for 12 point font.
%
    \gdef\twelvepointsgl{%
         \twelvepoint%
         \abovedisplayskip 14pt plus 4pt minus 4pt%
         \belowdisplayskip 14pt plus 4pt minus 4pt%
         \abovedisplayshortskip 0pt plus 4pt%
         \belowdisplayshortskip 8pt plus 4pt minus 4pt%
         \baselineskip 14pt%
         \lineskip      3pt%
         \lineskiplimit 1pt%
         \parindent     0pt%
         \parskip       0pt plus 1pt}%
%
   %
   %
%
%   This is a set of macros for switching to 12 point
%
\catcode`@=11\relax%
\def\definetwelve{%
   \font\twelvetext=cmr12 scaled 1000%
   \font\twelvescript=cmr9 scaled 1000%
   \font\twelvescriptscript=cmr7 scaled 1000%
   \font\twelvemathtext=cmmi12 scaled 1000%
   \font\twelvemathscript=cmmi9 scaled 1000%
   \font\twelvemathscriptscript=cmmi7 scaled 1000%
   \font\twelvesymboltext=cmsy10 scaled 1200%             was 1440
   \font\twelvesymbolscript=cmsy9 scaled 1000%
   \font\twelvesymbolscriptscript=cmsy7 scaled 1000%
   \font\twelvemathex=cmex10 scaled 1200%                 was 1440
   \font\twelveboldtext=cmbx12 scaled 1000%
   \font\twelveboldscript=cmbx9 scaled 1000%
   \font\twelveboldscriptscript=cmbx7 scaled 1000%
   \font\twelveslant=cmsl12 scaled 1000%
   \font\twelveitalic=cmti12 scaled 1000%
   \font\twelvetypewriter=cmtt12 scaled 1000%
% Set skew and hyphenation characters
   \skewchar\twelvemathtext='177%
   \skewchar\twelvemathscript='177%
   \skewchar\twelvemathscriptscript='177%
   \skewchar\twelvesymboltext='60%
   \skewchar\twelvesymbolscript='60%
   \skewchar\twelvesymbolscriptscript='60%
   \hyphenchar\twelvetypewriter=-1%
   \global\let\definetwelve=\relax}%
\def\twelvepoint{%
   \definetwelve%
   \textfont0=\twelvetext \scriptfont0=\twelvescript%
      \scriptscriptfont0=\twelvescriptscript%
   \textfont1=\twelvemathtext \scriptfont1=\twelvemathscript%
      \scriptscriptfont1=\twelvemathscriptscript%
   \textfont2=\twelvesymboltext \scriptfont2=\twelvesymbolscript%
      \scriptscriptfont2=\twelvesymbolscriptscript%
   \textfont3=\twelvemathex \scriptfont3=\twelvemathex%
      \scriptscriptfont3=\twelvemathex%
   \textfont\bffam=\twelveboldtext \scriptfont\bffam=\twelveboldscript%
      \scriptscriptfont\bffam=\twelveboldscriptscript%
   \textfont\itfam=\twelveitalic%
   \textfont\slfam=\twelveslant%
   \textfont\ttfam=\twelvetypewriter%
   \def\rm{\fam0\twelvetext}%
   \def\bf{\fam\bffam\twelveboldtext}%
   \def\it{\fam\itfam\twelveitalic}%
   \def\sl{\fam\slfam\twelveslant}%
   \def\tt{\fam\ttfam\twelvetypewriter}%
%
%   Set up \mit, \cal, \oldstyle 
%
   \def\mit{\fam1 }%
   \def\cal{\fam2 }%
   \def\oldstyle{\fam1 \twelvemathtextfont}%
%
%   Set up big etc.
%
\begingroup
\gdef\big##1{{\hbox{$ \left##1\vbox to13.8pt{}\right.\n@space$}}}%   10.2pt
\gdef\Big##1{{\hbox{$ \left##1\vbox to17.4pt{}\right.\n@space$}}}%   13.8pt
\gdef\bigg##1{{\hbox{$\left##1\vbox to21.0pt{}\right.\n@space$}}}%   17.4pt
\gdef\Bigg##1{{\hbox{$\left##1\vbox to24.6pt{}\right.\n@space$}}}%   21.0pt
\endgroup
%
%   Now define a strut box for twelve point.  Should be the height and depth
%   of a \Sigma in math mode.
%
   \setbox\strutbox=\hbox{\vrule height11.5pt depth3.5pt width0pt}%
%
%   Define baselineskip
%
   \normalbaselineskip=14pt%
   \normalbaselines\rm%
}
\catcode`@=12
\twelvepointdbl

%% file: 17ptdbl.tex
%
%
% Double space macro for 17 point font.
%
     \gdef\seventeenpointdbl{%
          \seventeenpoint%
          \abovedisplayskip 22pt plus 6pt minus 6pt%
          \belowdisplayskip 22pt plus 6pt minus 6pt%
          \abovedisplayshortskip  0pt plus 6pt%
          \belowdisplayshortskip 13pt plus 6pt minus 6pt%
          \baselineskip 27pt%
          \lineskip     11pt%
          \lineskiplimit 0pt%
          \parindent     0pt%
          \parskip       0pt plus 6pt}%
%
% Single space macro for 17 point font.
%
    \gdef\seventeenpointsgl{%
         \seventeenpoint%
         \abovedisplayskip 22pt plus 4pt minus 6pt%
         \belowdisplayskip 22pt plus 4pt minus 6pt%
         \abovedisplayshortskip 0pt plus 4pt%
         \belowdisplayshortskip 10pt plus 4pt minus 5pt%
         \baselineskip 19pt%
         \lineskip      2pt%
         \lineskiplimit 0pt%
         \parindent     0pt%
         \parskip       0pt plus 2pt}%
%
    %
    %
%
%   This is a set of macros for switching to 17 point
%
\catcode`@=11\relax%
\newfam\bfslfam%
\def\defineseventeen{%
   \font\seventeentext=cmr17 scaled 1000%
   \font\seventeenscript=cmr12 scaled 1000%
   \font\seventeenscriptscript=cmr9 scaled 1000%
   \font\seventeenmathtext=cmmi12 scaled 1440%
   \font\seventeenmathscript=cmmi12 scaled 1000%
   \font\seventeenmathscriptscript=cmmi9 scaled 1000%
   \font\seventeensymboltext=cmsy10 scaled 1728%
   \font\seventeensymbolscript=cmsy10 scaled 1440%
   \font\seventeensymbolscriptscript=cmsy9 scaled 1000%
   \font\seventeenmathex=cmex10 scaled 1728%
   \font\seventeenboldtext=cmbx12 scaled 1440%
   \font\seventeenboldscript=cmbx12 scaled 1000%
   \font\seventeenboldscriptscript=cmbx9 scaled 1000%
   \font\seventeenslant=cmsl12 scaled 1440%
%  \font\seventeenboldslant=cmbxsl10 scaled 1728%   what about script etc?
   \font\seventeenboldslant=cmbxsl10 scaled 1440%   the best I can do...
   \font\seventeenitalic=cmti12 scaled 1440%
   \font\seventeentypewriter=cmtt12 scaled 1440%
% Set skew and hyphenation characters  -- why not all fonts?
   \skewchar\seventeenmathtext='177%
   \skewchar\seventeenmathscript='177%
   \skewchar\seventeenmathscriptscript='177%
   \skewchar\seventeensymboltext='60%
   \skewchar\seventeensymbolscript='60%
   \skewchar\seventeensymbolscriptscript='60%
   \hyphenchar\seventeentypewriter=-1%
   \global\let\defineseventeen=\relax} % avoids having to re-define fonts
                                       % with evry change to 17 point.
\def\seventeenpoint{%
   \defineseventeen%
   \textfont0=\seventeentext \scriptfont0=\seventeenscript%
      \scriptscriptfont0=\seventeenscriptscript%
   \textfont1=\seventeenmathtext \scriptfont1=\seventeenmathscript%
      \scriptscriptfont1=\seventeenmathscriptscript%
   \textfont2=\seventeensymboltext \scriptfont2=\seventeensymbolscript%
      \scriptscriptfont2=\seventeensymbolscriptscript%
   \textfont3=\seventeenmathex \scriptfont3=\seventeenmathex%
      \scriptscriptfont3=\seventeenmathex%
   \textfont\bffam=\seventeenboldtext \scriptfont\bffam=\seventeenboldscript%
      \scriptscriptfont\bffam=\seventeenboldscriptscript%
   \textfont\itfam=\seventeenitalic%
   \textfont\slfam=\seventeenslant%
   \textfont\ttfam=\seventeentypewriter%
   \def\rm{\fam0\seventeentext}%
   \def\bf{\fam\bffam\seventeenboldtext}%
   \def\it{\fam\itfam\seventeenitalic}%
   \def\sl{\fam\slfam\seventeenslant}%
   \def\tt{\fam\ttfam\seventeentypewriter}%
%
% Creating a new family : bold face slant.
%
% I haven't included any script or scriptscript fonts or the appropriate
% skew and hypenation characters.
%
%  \newfam\bfslfam%  must be defined outside of this macro ie. executed once.
   \textfont\bfslfam=\seventeenboldslant%
   \def\bfsl{\fam\bfslfam\seventeenboldslant}%
   \def\slbf{\fam\bfslfam\seventeenboldslant}%
%
%   Set up \mit, \cal, \oldstyle 
%
   \def\mit{\fam1 }%
   \def\cal{\fam2 }%
   \def\oldstyle{\fam1 \seventeenmathtextfont}%
%
%   Set up big etc.
%
\begingroup
\gdef\big##1{{\hbox{$ \left##1\vbox to17.4pt{}\right.\n@space$}}}%   13.8/10.2pt
\gdef\Big##1{{\hbox{$ \left##1\vbox to21.0pt{}\right.\n@space$}}}%   17.4/13.8pt
\gdef\bigg##1{{\hbox{$\left##1\vbox to24.6pt{}\right.\n@space$}}}%   21.0/17.4pt
\gdef\Bigg##1{{\hbox{$\left##1\vbox to28.2pt{}\right.\n@space$}}}%   24.6/21.0pt
\endgroup
%
%   Now define a strut box for seventeen point.  Should be obtained as the
%   height and depth for \Sigma.
%
   \setbox\strutbox=\hbox{\vrule height21.5pt depth3.5pt width0pt}%
%
%  This definition used by SliTeX.
%
%  \setbox\strutbox=\hbox{\vrule height13.5pt depth5.6pt width0pt}
%
%   Define baselineskip, must agree with definition above.
%
   \normalbaselineskip=19pt%
   \normalbaselines\rm%
}
\catcode`@=12
\seventeenpointdbl

%% file: tilde.tex
% the \lower command is used to redfine the reference point for ~
% this lowered ~ is placed in a box with zero height and depth.
% the net effect is to ensure that we don't create extra white space
% below the ~.
% ---------------------------------------------------------------------------
%
\newdimen\LCBtmp%
\def\maketildevec#1#2{%
   \vtop{\baselineskip=0pt\lineskip=0pt%
   \setbox1=\hbox{$\seventeenpoint#1\char`\~$}%
   \LCBtmp=1.2\ht1%
   \setbox0=\hbox{\lower\LCBtmp\hbox{\box1}}\ht0=0pt\dp0=0pt%
   \ialign{\hfil ##\hfil\crcr\hbox{$#1#2$}\crcr\box0\crcr}}}%
%
% As above, but places a larger \tilde above a character than normally used
% Plain TeX
%
\def\Tilde#1{\mathchoice%
   {\maketilde{\textstyle}{#1}}%
   {\maketilde{\displaystyle}{#1}}%
   {\maketilde{\scriptstyle}{#1}}%
   {\maketilde{\scriptscriptstyle}{#1}}}%
\def\maketilde#1#2{%
   \vbox{\baselineskip=0pt\lineskip=0pt%
   \setbox1=\hbox{$\seventeenpoint#1\char`\~$}%
   \LCBtmp=0.6\ht1%
   \setbox0=\hbox{\lower\LCBtmp\hbox{\box1}}\ht0=0pt\dp0=0pt%
   \ialign{\hfil##\hfil\crcr\box0\crcr\hbox{$#1#2$}\crcr}}}%

%% file: rotate.tex
%
%   These macros allow you to rotate or flip a \TeX\ box.  Very useful for
%   sideways tables or upsidedown answers.
%
%   To use, create a box containing the information you want to rotate.
%   (An hbox or vbox will do.)  Now call \rotr\boxnum to rotate the
%   material and create a new box with the appropriate (flipped) dimensions.
%   \rotr rotates right, \rotl rotates left, \rotu turns upside down, and
%   \rotf flips.  These boxes may contain other rotated boxes.
%
\newdimen\rotdimen
\def\vspec#1{\special{ps:#1}}%  passes #1 verbatim to the output
\def\rotstart#1{\vspec{gsave currentpoint currentpoint translate
   #1 neg exch neg exch translate}}% #1 can be any origin-fixing transformation
\def\rotfinish{\vspec{currentpoint grestore moveto}}% gets back in synch
%
%   First, the rotation right. The reference point of the rotated box
%   is the lower right corner of the original box.
%
\def\rotr#1{\rotdimen=\ht#1\advance\rotdimen by\dp#1%
   \hbox to\rotdimen{\hskip\ht#1\vbox to\wd#1{\rotstart{90 rotate}%
   \box#1\vss}\hss}\rotfinish}
%
%   Next, the rotation left. The reference point of the rotated box
%   is the upper left corner of the original box.
%
\def\rotl#1{\rotdimen=\ht#1\advance\rotdimen by\dp#1%
   \hbox to\rotdimen{\vbox to\wd#1{\vskip\wd#1\rotstart{270 rotate}%
   \box#1\vss}\hss}\rotfinish}%
%
%   Upside down is simple. The reference point of the rotated box
%   is the upper right corner of the original box. (The box's height
%   should be the current font's xheight, \fontdimen5\font,
%   if you want that xheight to be at the baseline after rotation.)
%
\def\rotu#1{\rotdimen=\ht#1\advance\rotdimen by\dp#1%
   \hbox to\wd#1{\hskip\wd#1\vbox to\rotdimen{\vskip\rotdimen
   \rotstart{-1 dup scale}\box#1\vss}\hss}\rotfinish}%
%
%   And flipped end for end is pretty ysae too. We retain the baseline.
%
\def\rotf#1{\hbox to\wd#1{\hskip\wd#1\rotstart{-1 1 scale}%
   \box#1\hss}\rotfinish}%